\documentclass[journal,twoside]{IEEEtran}
\IEEEoverridecommandlockouts                              % This command is only
\usepackage{epsfig} % for postscript graphics files
\usepackage{xcolor}
\usepackage{eurosym}
\usepackage{amsmath,amsfonts,amssymb}
\usepackage{cite}
\usepackage{color}
\usepackage{graphicx}
\usepackage{array}
\usepackage{textcomp}
\usepackage{siunitx}

\newcommand{\qed}{\hspace*{\fill}$\Box$\par}
\newtheorem{theorem}{Theorem}

\newtheorem{definition}{Definition}

\newtheorem{allocation}{Allocation}

\title{%
Sharing Storage in a Smart Grid: A Coalitional Game Approach%
}

\author{%
Pratyush Chakraborty$^*$,
Enrique Baeyens$^*$,
Kameshwar Poolla,
Pramod P. Khargonekar, and
Pravin Varaiya
\thanks{This research is supported by the National Science Foundation under grants EAGER-1549945, CPS-1646612, CNS-1723856 and by the National Research Foundation of Singapore under a grant to the Berkeley Alliance for Research in Singapore}% <-this % stops a space
\thanks{$^*$The first two authors contribute equally}
\thanks{Corresponding author P.~Chakraborty is with the
Department of Mechanical Engineering, University of California, Berkeley, CA, USA}%
\thanks{E.~Baeyens is with the Instituto de las Tecnolog\'{\i}as Avanzadas de la Producci\'on,
Universidad de Valladolid, Valladolid, Spain}%
\thanks{K. Poolla and P. Varaiya are with the Department of Electrical Engineering and Computer Science, University of California, Berkeley, CA, USA}
\thanks{P.~P.~Khargonekar is with the Department of Electrical Engineering and Computer Science, University of California, Irvine, CA, USA}
}

\begin{document}

\maketitle

%%%%%%%%%%%%%%%%%%%%%%%%%%%%%%%%%%%%%%%%%%%%%%%%%%%%%%%%%%%%%%%%%%%%%%%%%%%%%%%%
\begin{abstract}
Sharing economy is a transformative socio-economic phenomenon built around the idea of sharing underused resources and services, \emph{e.g.} transportation and housing, thereby reducing costs and extracting value. Anticipating continued reduction in the cost  of electricity storage, we look into the potential opportunity in electrical
power system where consumers share storage with each other.
We consider two different scenarios. In the first scenario, consumers are assumed to
already have individual storage devices and they explore cooperation to minimize the
realized electricity consumption cost. In the second scenario, a group of
consumers is interested to invest in joint storage capacity and operate it
cooperatively. The resulting system problems are modeled using cooperative game theory.
 In both cases, the cooperative games are shown to have
non-empty cores and we develop efficient cost allocations in the core
with analytical expressions.  Thus, sharing of storage in cooperative manner
is shown to be very effective for the electric power system.
\end{abstract}

\begin{IEEEkeywords}
Storage Sharing, Cooperative Game Theory, Cost Allocation
\end{IEEEkeywords}
%%%%%%%%%%%%%%%%%%%%%%%%%%%%%%%%%%%%%%%%%%%%%%%%%%%%%%%%%%%%%%%%%%%%%%%%%%%%%%%%

\section{Introduction}

\subsection{Motivation}

The sharing economy is disruptive and transformative socio-economic trend that has already impacted transportation and housing \cite{heinrichs2013}. People rent out (rooms in) their houses
and use their cars to provide transportation services. The business model of sharing economy leverages under
utilized resources.  Like these sectors, many
of the resources in electricity grid is also under-utilized or under-exploited.  
There is potential benefit in sharing the excess generation by rooftop
solar panels, sharing flexible demand, sharing unused
capacity in the storage services, etc. Motivated by the recent studies
\cite{Kittner2017} predicting a fast drop in battery storage prices, we focus on
sharing electric energy storage among consumers.

\subsection{Literature Review}
Storage prices are projected to decrease by more than $30\%$ by 2020.
The arbitrage value and welfare effects of storage in electricity markets has
been explored in literature. In \cite{graves1999}, the value of storage
arbitrage was studied in deregulated markets. In \cite{sioshansi2009}, the
authors studied the role of storage in wholesale electricity markets. The
economic viability of the storage elements through price arbitrage was examined
in \cite{bradbury2014}. Agent-based models to explore the tariff arbitrage
opportunities for residential storage systems were introduced in
\cite{zheng2014}.  
In \cite{wu2017,van2013}, authors address the optimal control and
coordination of energy storage. All these works explore the economic value of
storage to an individual, not for shared services. Sharing of storage among
firms has been analyzed using non-cooperative game theory in~\cite{wu2016}.
But the framework needs a spot market among the consumers and also coordination
is needed among the firms that are originally strategic.

In this paper, we explore sharing storage in a cooperative manner among
consumers.  Cooperative game theory has significant potential to model resource
sharing effectively \cite{Saad2009}. Cooperation and aggregation of renewable
energy sources bidding in a two settlement market to maximize expected and
realized profit has been analyzed using cooperative game
theory in~\cite{chakraborty2016, baeyens2013, chakraborty2016phd}. Under a
cooperative set-up, the cost allocation to all the agents is a crucial task. A
framework for allocating cost in a fair and stable way was introduced
in~\cite{chakraborty2017}. Cooperative game theoretic analysis of multiple
demand response aggregators in a virtual power plant and their cost allocation
has been tackled in~\cite{nguyen2017}. In \cite{chakraborty20172}, sharing
opportunities of photovoltaic systems (PV) under various billing mechanisms
were explored using cooperative game theory.  

\subsection{Contributions and Paper Organization} 
In this paper, we investigate the sharing of storage systems in a time of use
(TOU) price set-up using cooperative game theory. We consider two scenarios. In
the first one, a group of consumers already own storage systems and they are
willing to operate all together to minimize their electricity consumption cost.
In a second scenario, a group of consumers wish to invest in a shared common
storage system and get benefit for long term operation in a cooperative manner.
We model both the cases using cooperative game theory.  We prove that the
resulting games developed have non-empty cores, \emph{i.e.,} cooperation is
shown to be beneficial in both the cases.  We also derive closed-form and easy
to compute expressions for cost allocations in the core in both the cases. Our
results suggest that sharing of electricity storage in a cooperative manner is
an effective way to amortize storage costs and to increase its utilization. In
addition, it can be very much helpful for consumers and at the same time to
integrate renewables in the system, because off-peak periods correspond to
large presence of renewables that can be stored for consumption during peak
periods.

The remainder of the paper is organized as follows. In Section~\ref{sec:prob},
we formulate the cooperative storage problems.
A brief review of cooperative game theory is
presented in Section \ref{sec:coalg}. In Section~\ref{sec:mainr}, we state and
explain our main results. A case study illustrating our results
using real data from Pecan St. Project is presented in Section~\ref{sec:case}.
Finally, we conclude the paper in Section \ref{sec:concl}.

\section{Problem Formulation}
\label{sec:prob}
\subsection{System Model}

\begin{figure}
\centering
  \includegraphics[width=.6\columnwidth]{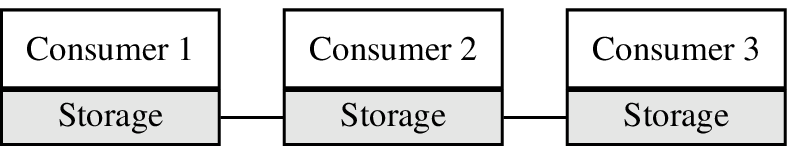} \\[4ex]
  \includegraphics[width=.6\columnwidth]{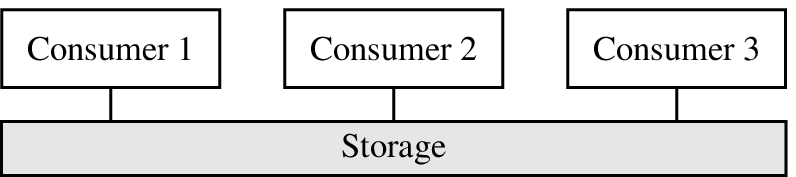}
  \caption{Configuration of three consumers in the two analyzed scenarios}
\label{fig:1}
\end{figure}

We consider a set of consumers indexed by $i \in \mathcal N:=\{1,2,\ldots,N\}$.
The consumers invest in storage. The consumers cooperate and share
their storage with each other. We consider two scenarios here.
In the scenario I, the consumers already have storage and they operate with
storage devices connected to each other. In the scenario II, the consumers
wish to invest in a common storage. There is a single meter for this group of
consumers. We assume that there is necessary electrical connection
between all the consumers for effective sharing. We ignore here the capacity
constraints, topology or losses in the connecting network. The configuration of
the scenarios with three consumers are depicted in Figure~\ref{fig:1}.
Examples of the situations considered here include consumers in an industrial
park, office buildings on a campus, or homes in a residential complex.

\subsection{Cost of Storage}

Each day is divided into two periods --peak and off-peak. There is a time-of-use pricing. The peak and off-peak period prices are denoted by $\pi_h$ and $\pi_{\ell}$ respectively. The prices are fixed and known to all the consumers.

Let $\pi_i$ be the daily capital cost of storage of the consumer
$i\in\mathcal N$ amortized over its life span.
Let the arbitrage price be defined by
\begin{equation}
\pi_\delta:=\pi_h-\pi_{\ell}
\end{equation}
and define the arbitrage constant $\gamma_i$ as follows:
\begin{equation}
\gamma_i:=\frac{\pi_{\delta}-\pi_i}{\pi_\delta}
\end{equation}
In order to have a viable arbitrage opportunity, we need
\begin{equation}
\pi_i \leq \pi_\delta
\end{equation}
which corresponds to $\gamma_i \in [0,1]$.
The consumers discharge their storage during peak hours and charge them during
off-peak hours.

The daily cost of storage of a consumer $i\in\mathcal N$ for the peak period
consumption $\mathbf x_i$ depends on the capacity investment $C_i$
and is given by
\begin{align}\label{eq:rcost}
J(\mathbf x_i,C_i)=\pi_i C_i + \pi_h (\mathbf x_i-C_i)^{+} +
\pi_{\ell} \min\{C_i,\mathbf x_i\},
\end{align}
where
$\pi_i C_i$ is the capital cost of acquiring $C_i$ units of storage capacity,
$\pi_h (\mathbf x_i-C_i)^{+}$ is the daily cost of the electricity purchase
during peak price period, and
$\pi_{\ell} \min\{\mathbf x_i,C_i\}$ is the daily cost of the
electricity purchase during off-peak period to be stored for
consumption during the peak period.
We ignore the off-peak period electricity consumption of the consumer from the
expression of $J$ as its expression is independent of the storage capacity.
%In determining the expression of the daily cost, we consider
%that the consuner only needs to purchase at hight
%price $\pi_h$ the difference between its consumption during the peak
%period and the stored electricity during the off-peak period.
%The off-peak consumption does not play any role in the
%cost of storage capacity. The reason is that off-peak consumption
%is charged at the lowest price $\pi_{\ell}$ at which electricity is available
%and storage capacity does not affects the cost of this consumption.
The daily peak consumption of electricity is not known in advance and
we assume it to be a random variable. Let $F$ be the joint cumulative
distribution function (CDF) of the collection of random variables
$\{\mathbf x_i : i\in\mathcal N\}$ that represents the consumptions of the
consumers in $\mathcal N$. If $\mathcal S \subseteq \mathcal N$ is a subset
of consumers, then $\mathbf x_{\mathcal S}$ denotes the aggregated peak
consumption of $\mathcal S$ and its CDF is $F_{\mathcal S}$.

The daily cost of storage of a group of consumers $\mathcal S \subseteq
\mathcal N$ with aggregated peak consumption
$\mathbf x_{\mathcal S}=\sum_{i\in\mathcal S}\mathbf x_i$
and joint storage capacity
$C_{\mathcal S}$ is
\begin{align}\label{eq:scost}
J(\mathbf x_{\mathcal S},C_{\mathcal S})=
\pi_{\mathcal S} C_{\mathcal S} +
\pi_h (\mathbf x_{\mathcal S}-C_{\mathcal S})^{+} +
\pi_{\ell} \min\{C_{\mathcal S},\mathbf x_{\mathcal S}\}
\end{align}
where $\pi_S$ is the daily capital cost of aggregated storage of the group
amortized during its life span. Note that the individual storage costs
(\ref{eq:rcost}) are obtained from (\ref{eq:scost}) for the singleton
sets $\mathcal S=\{i\}$.

The daily cost of storage given by (\ref{eq:rcost}) and (\ref{eq:scost})
are random variables with expected values
\begin{align}
J_{\mathcal S}(C_{\mathcal S}) =
\mathbb E J_{\mathcal S}(\mathbf x_{\mathcal S},C_{\mathcal S}),
\quad \mathcal S \subseteq \mathcal N.
\end{align}
In the sequel, we will distinguish between the random variables
and their realized values by using bold face fonts
$\mathbf x_{\mathcal S}$ for the random variables and
normal fonts  $x_{\mathcal S}$ for their realized values.

\subsection{Quantifying the Benefit of Cooperation Benefit}

We are interested in studying and quantifying the benefit of cooperation in the two scenarios.
In the first scenario, the consumers already have installed storage capacity
$\{C_i:i\in\mathcal N\}$ that
they acquired in the past. Each of the consumers can have a different
storage technology that was acquired at a different time compared to the other
consumers. Consequently, each consumer has a different daily capital cost
$\pi_i$. The consumers aggregate their storage capacities and they operate
using the same strategy, they use the aggregated storage capacity to
store energy during off-peak periods that they will later use during
peak periods. By aggregating storage devices, the unused capacity
of some consumers is used by others producing cost savings for
the group. We analyze this scenario using cooperative game theory and
develop an efficient allocation rule of the daily storage cost that
is satisfactory for every consumer.

In the second scenario, we consider a group of consumers that join to
buy storage capacity that they want to use in a cooperative way.
First, the group of consumers have to make a decision about how much
storage capacity they need to acquire and then they have to share the
expected cost among the group participants. The decision  problem
is modeled as an optimization problem where the group of consumers minimize the
expected cost of daily storage. The problem of sharing the expected
cost is modeled using cooperative game theory.
We quantify the reduction in the expected cost of storage for the
group and  develop a mechanism to allocate the expected cost
among the participants that is satisfactory for all of them.

\section{Background: Coalitional Game Theory for Cost Sharing}
\label{sec:coalg}

Game theory deals with rational behavior of economic agents in a mutually
interactive setting \cite{Neumann1944}. Broadly speaking, there are two major categories of games: non-cooperative games and cooperative games.
Cooperative games (or coalitional games) have been used extensively in diverse
disciplines such as social science, economics, philosophy,
psychology and communication networks \cite{Myerson2013,Saad2009}. Here, we focus on cooperative games for cost sharing \cite{Jain2007}.

Let $\mathcal{N}:=\{1,2,\ldots,N\}$ denote a finite collection of players. In a cooperative game for cost sharing, the players want to minimize their joint cost and share the resulting cost cooperatively.
\begin{definition}[Coalition]
A \emph{coalition} is any subset $\mathcal S \subseteq \mathcal N$.
The number of players in a coalition $\mathcal S$ is denoted by its cardinality, $|\mathcal S|$.
The \emph{set of all possible coalitions} is defined as the power set $2^{\mathcal N}$ of $\mathcal N$.
The \emph{grand coalition} is the set of all players, $\mathcal N$.
\end{definition}

\begin{definition}[Game and Value]
A cooperative game is defined by a pair $(\mathcal{N},v)$ where
$v:2^{\mathcal{N}}\rightarrow \mathbb R$
is the \emph{value function} that assigns a real value to each coalition
$\mathcal{S} \subseteq \mathcal{N}$.
Hence, the \emph{value of coalition} $\mathcal S$ is given by $v(\mathcal S)$. For the cost sharing game, $v(\mathcal S)$ is the total cost of the coalition.
\end{definition}

\begin{definition}[Subadditive Game]
A cooperative game $(\mathcal N,v)$ is \emph{subadditive} if, for any pair of disjoint coalitions
$\mathcal S, \mathcal T \subset \mathcal N$ with
$\mathcal S \cap \mathcal T = \emptyset$, we have $v(\mathcal S) + v(\mathcal T) \ \geq \ v(\mathcal S \cup  \mathcal T).$
\end{definition}

Here we consider the value of the coalition $v(\mathcal S)$ is \emph{transferable} among players.
The central question for a subadditive cost sharing game with transferrable value is how  to \emph{fairly} distribute the coalition value among the coalition members.

\begin{definition}[Cost Allocation]
A \emph{cost allocation} for the coalition $\mathcal{S} \subseteq \mathcal N$
is a vector $x\in\mathbb{R}^{N}$
whose entry $x_i$ represents the allocation to member $i \in \mathcal S$
($x_i = 0, \ \ i \notin \mathcal S$).
\end{definition}
For any coalition $\mathcal S\subseteq{\mathcal N}$, let $x_{\mathcal S}$
denote the sum of cost allocations for every coalition member, \emph{i.e.}
$x_{\mathcal S}=\sum_{i\in\mathcal S}x_i$.

%\begin{definition}[Dissatisfaction and excess] \label{def:excess}
%The \emph{dissatisfaction} of  a coalition $\mathcal S$ with respect to the
%payoff allocation $x$ is measured by the {\em excess} defined as follows:
%\begin{equation}
%e(x,\mathcal S)=v(\mathcal S)-\sum_{i\in \mathcal S}x_i.
%\end{equation}
%\end{definition}
%An allocation is called {\em efficient} if its excess is zero.
\begin{definition}[Imputation]
A cost allocation $x$ for the grand coalition $\mathcal N$ is said to be an
{\em imputation} if it is simultaneously efficient
--\emph{i.e.} $v(\mathcal N)=x_{\mathcal N}$, and individually rational
--\emph{i.e.} $v({i})\geq x_i, \forall i \in \mathcal N$.
Let $\mathcal I$ denote the set of all imputations.
\end{definition}

The fundamental solution concept for cooperative games is the \emph{core}
\cite{Neumann1944}.

\begin{definition}[The Core]  \label{core_def}
The \emph{core} $\mathcal{C}$ for the cooperative game $(\mathcal{N},v)$
with \emph{transferable cost} is defined as the set of cost allocations
such that no coalition can have cost which is lower than the
sum of the members current costs under the given allocation.
\begin{align}
\label{core_eq}
\mathcal C :=
\left\{
x \in \mathcal I:
v(\mathcal S)\geq x_{\mathcal S},
\forall \mathcal S \in 2^{\mathcal N}
\right\}.
\end{align}
\end{definition}

%Games can have empty cores. Let us first define a \emph{concave game}.

%\begin{theorem}[Concave Game\cite{Shapley1971}] A cooperative game has a
%\emph{nonempty core} if it is \emph{concave}
%--\emph{i.e.}, has a \emph{submodular} value function,
%\begin{equation}
%\label{eq:supermod1}
%v(\mathcal S) + v(\mathcal T)  \ \geq \ v(\mathcal S \cup \mathcal T) + v(\mathcal S \cap \mathcal T), \ \ \text{for all} \ \ \mathcal %S,\mathcal T \subset \mathcal N
%\end{equation}
%\end{theorem}
%Many real world cost sharing games are not concave.
A classical result in cooperative game theory, known as Bondareva-Shapley theorem, gives a necessary and sufficient condition for a game to have nonempty core. To state this theorem, we need the following definition.

\begin{definition}[Balanced Game and Balanced Map]
A cooperative game $(\mathcal N, v)$ for cost sharing is \emph{balanced} if for any balanced map
$\alpha$,
$\sum_{\mathcal S\in 2^{\mathcal N}}
\alpha(\mathcal S) v(\mathcal S) \geq v(\mathcal N)$ where the map
$\alpha: 2^{\mathcal N} \rightarrow [0,1]$ is said to be \emph{balanced}
if for all $i\in\mathcal N$, we have $\sum_{\mathcal S\in 2^{\mathcal N}}
\alpha(\mathcal S) \mathbf{1}_{\mathcal S}(i) = 1$, where
$\mathbf{1}_{\mathcal S}$ is the indicator function of the set $\mathcal S$,
\emph{i.e.} $\mathbf{1}_{\mathcal S}(i)=1$ if $i\in\mathcal S$ and
$\mathbf{1}_{\mathcal S}(i)=0$ if $i\not\in\mathcal S$.
\end{definition}
Next we state the Bondareva-Shapley theorem.

\begin{theorem}[Bondareva-Shapley Theorem \cite{Saad2009}]
A coalitional game has a nonempty core if and only if it is balanced.
\end{theorem}

%There are a number of promising solutions that exist for a cooperative game.
%The most prominent of them are the \emph{Shapley value} and the \emph{nucleolus} \cite{Myerson2013}.
If a game is balanced, the nucleolus  \cite{Myerson2013} is a solution that is always in the core.
% For a concave cost sharing game, the Shapley value is in the core.

\section{Main Results}
\label{sec:mainr}
\subsection{Scenario I: Realized Cost Minimization with Already Procured Storage Elements}
Our first concern is to study if there is some benefit in cooperation of the
consumers by sharing the storage capacity that they already have. To analyze
this scenario  we shall formulate our problem as a coalitional game.
\subsubsection{Coalitional Game and Its Properties}
The players of the cooperative game are
the consumers that share their storage and want to reduce their realized
joint storage investment cost. For any coalition
$\mathcal S \subseteq \mathcal N$, the cost of the coalition is
$u(\mathcal S)$ which is the realized cost of the joint
storage investment $C_{\mathcal S}=\sum_{i\in\mathcal S}C_i$. Each
consumer may have a different daily capital cost of storage
$\{\pi_{i}:i\in\mathcal N\}$, because they
did not necessarily their storage systems at the same time or at the same
price for KW. The realized cost of the joint storage for the peak
period consumption $x_{\mathcal S}=\sum_{i \in \mathcal S} x_i$
is given by
\begin{align}
u(\mathcal S)&=J(x_{\mathcal S},C_{\mathcal S})
%\nonumber\\
%&=\sum_{i\in\mathcal S}\pi_i C_i + \pi_h (x_{\mathcal S}-C_{\mathcal S})^{+}
%  +\pi_{\ell} \min\{C_{\mathcal S},x_{\mathcal S}\},
\label{eq:valdef}
\end{align}
where $J$ was defined in (\ref{eq:scost}). Since we are using the
realized value of the aggregated peak consumption $x_{\mathcal S}$,
$J(x_{\mathcal S},C_{\mathcal S})$ is not longer a random variable.

In order to show that cooperation is advantageous for the members
of the group, we have to prove that the game is subadditive.
In such a case, the joint daily investment cost of the consumers
is never greater that the sum of the individual daily investment costs.
Subadditivity of the cost sharing coalitional game is established in
Theorem~\ref{th:subadd1}.

\begin{theorem}\label{th:subadd1}
The cooperative game for storage investment cost sharing
$(\mathcal N, u)$ with the cost function
$u$ defined in (\ref{eq:valdef}) is subadditive.
\end{theorem}
\paragraph*{Proof} See appendix.

However, subadditivity is not enough to provide satisfaction of the
coalition members. We need a stabilizing allocation mechanism of the
aggregated cost. Under a stabilizing cost sharing mechanism no member in
the coalition is impelled to break up the coalition. Such a mechanism
exists if the cost sharing coalitional game is
balanced. Balancedness of the cost sharing coalitional game is
established in Theorem~\ref{th:bal1}.

\begin{theorem} \label{th:bal1}
The cooperative game for storage investment cost sharing
$(\mathcal N, u)$ with the cost function $u$
defined in (\ref{eq:valdef}) is balanced.
\end{theorem}
\paragraph*{Proof} See the appendix.
\subsubsection{Sharing of Realized Cost}
Since the cost sharing cooperative game $(\mathcal N, u)$ is balanced, its core
is nonempty and there always exist cost allocations that stabilize
the grand coalition. One of this coalitions is the nucleolus while another one is
the allocation that minimizes the worst case excess \cite{baeyens2013}.
However, computing these allocations requires solving linear programs
with a number of constraints that grows exponentially with the cardinality of
the grand coalition and they can be only applied for coalitions of moderate
size. As an alternative to these computationally intensive cost allocations,
we propose the following cost allocation.

\begin{allocation} \label{alloc:1}
Define the cost allocation $\{\xi_i : i\in\mathcal N\}$ as follows:
\begin{align}
\xi_i :=
\left\{
\begin{array}{ll}
\pi_i C_i+ \pi_h(x_i-C_i)+\pi_{\ell} C_i, &
\mathrm{if} \ x_{\mathcal N} \geq C_{\mathcal N}\\
\pi_i C_i + \pi_{\ell} x_i, &
\mathrm{if} \ x_{\mathcal N} < C_{\mathcal N}
\end{array}
\right.
\label{eq:alloc1}
\end{align}
for all $i\in\mathcal N$.
\end{allocation}

We establish in Theorem~\ref{th:alloc1}, this cost allocation belongs to the core
of the cost sharing cooperative game.

\begin{theorem} \label{th:alloc1}
The cost allocation $\{\xi_i: i\in\mathcal N\}$ defined in
Allocation~\ref{alloc:1} belongs to the core of the cost sharing
cooperative game $(\mathcal N,u)$.
\end{theorem}
\paragraph*{Proof} See appendix.

Unlike the nucleolus or the cost allocation minimizing the worst-case
excess, Allocation 1 has an analytical expression and can be easily
obtained without any costly computation. Thus, we have developed a strategy
such that consumers that independently invested in storage, and are subject to a
two period (peak and off-peak)
TOU pricing mechanism can reduce their costs by sharing their storage
devices. Moreover, we have proposed a cost sharing allocation rule that
stabilizes the grand coalition. This strategy can be considered a weak
cooperation because each consumer acquired its storage capacity independently
of each other, but they agree to share the joint storage capacity.

In the next section we consider a stronger cooperation problem, where
a group of consumers decide to invest jointly in storage capacity.

\subsection{Scenario II: Expected Cost Minimization for Joint Storage Investment}

In this scenario, we consider a group of consumers indexed by $i\in\mathcal N$,
that decide to jointly invest in storage capacity. We are interested
in studying whether cooperation provides a benefit for the coalition members for the long term.

\subsubsection{Coalitional Game and Its Properties}
Similar to the previous case, only the peak consumption is relevant
in the investment decision. Let $\mathbf x_i$ denote the daily peak period
consumption of consumer $i\in\mathcal N$. Unlike the previous scenario, here
$\mathbf x_i$ is a random variable with marginal
cumulative distribution function (CDF) $F_i$.
The daily cost of the consumer $i\in\mathcal N$ depends on the storage
capacity investment of the consumer as per (\ref{eq:rcost}).
This cost is also a random variable. If the consumer is risk neutral, it acquires the storage
capacity $C_i^*$ that minimizes the expected value of the daily cost
\begin{align}
C_i^* &= \arg\min_{C_i\geq 0} J_i(C_i),
\end{align}
where
\begin{align}
J_i(C_i)&=\mathbb{E}J(\mathbf x_i,C_i),
%\nonumber\\
%&=\pi_{\mathcal S} C_i + \pi_h \mathbb{E}(\mathbf x_i-C_i)^{+} +
%\pi_{\ell} \mathbb{E}\min\{C_i,\mathbf x_i\}
\end{align}
and $\pi_{\mathcal S}$ is the daily capital cost of storage amortized over its
lifespan that in this case is the same for each of the consumers
--\emph{i.e.} $\pi_i=\pi_{\mathcal S}$ for all $i\in\mathcal N$, because
we assume that they buy storage devices of the same technology at the
same time.
This problem has been previously solved in \cite{wu2016} and its
solution is given by Theorem~\ref{th:opt2}.
\begin{theorem}[\cite{wu2016}]\label{th:opt2}
The storage capacity of a consumer $i\in\mathcal N$ that minimizes its
daily expected cost is $C^*_i$, where
\begin{align*}
F_i(C^*_i) = \frac{\pi_{\delta}-\pi_{\mathcal S}}{\pi_{\delta}}=\gamma_{\mathcal S}
\end{align*}
and the resulting optimal cost is
\begin{align}
J^*_i = J_i(C^*_i)
= \pi_{\ell} \mathbb{E}[\mathbf x_i] +
\pi_{\mathcal S} \mathbb{E}[\mathbf x_i\mid \mathbf x_i\geq C_i^*].
\end{align}
\end{theorem}

Let us consider a group of consumers $\mathcal S \subseteq N$ that decide
to join to invest in joint storage capacity. The joint peak consumption
of the coalition is
$\mathbf x_{\mathcal S}=\sum_{i\in\mathcal S}\mathbf x_i$ with
CDF $F_{\mathcal S}$.  We also assume that the joint
CDF of all the agent's peak consumptions $F$ is known or can be estimated
from historical data. By applying Theorem~\ref{th:opt2}, the optimal
investment in storage capacity of the coalition
$\mathcal S \subseteq \mathcal N$ is $C_{\mathcal S}^*$ such that
$F_{\mathcal S}(C_{\mathcal S}^*) = \gamma_{\mathcal S}$ and the optimal cost is
\begin{align}\label{eq:jsopt}
J_{\mathcal S}^* = J_{\mathcal S}(C_{\mathcal S}^*)
= \pi_{\ell} \mathbb{E}[\mathbf x_{\mathcal S}] +
  \pi_{\mathcal S} \mathbb{E}[\mathbf x_{\mathcal S} \mid \mathbf x_{\mathcal S}\geq C_{\mathcal S}^*].
\end{align}

Consider the cost sharing cooperative game $(\mathcal N, v)$ where the cost
function $v:2^{\mathcal N}\rightarrow\mathbb R$ is defined as follows
\begin{align} \label{eq:valdef2}
v(\mathcal S)
= J_{\mathcal S}^*
= \arg\min_{C_{\mathcal S}\geq 0} J_{\mathcal S}(C_{\mathcal S}),
\end{align}
where $J_{\mathcal S}^*$ was defined in (\ref{eq:jsopt}).

Similar to the case of consumers that already own storage capacity and
decide to join to reduce their costs, here we prove that the
cooperative game is subadditive so that the consumer obtain a reduction
of cost. This is the result in Theorem~\ref{th:subadd2}.

\begin{theorem}\label{th:subadd2}
The cooperative game for storage investment cost sharing
$(\mathcal N, v)$ with the cost function
$v$ defined in (\ref{eq:valdef2}) is subadditive.
\end{theorem}
\paragraph*{Proof} See appendix.

We also need a cost allocation rule that is stabilizing. Theorem~\ref{th:bal2} establishes that
the game is balanced and has a stabilizing allocation.

\begin{theorem}\label{th:bal2}
The cooperative game for storage investment cost sharing
$(\mathcal N, v)$ with the cost function
$v$ defined in (\ref{eq:valdef2}) is balanced.
\end{theorem}
\paragraph*{Proof} See appendix.

\subsubsection{Stable Sharing of Expected Cost}

Similar to the previous scenario, we were able to develop a
cost allocation rule that is in the core. This cost allocation rule
has an analytical formula and can be efficiently computed. This
allocation rule is defined as follows.

\begin{allocation} \label{alloc:2}
Define the cost allocation $\{\zeta_i : i\in\mathcal N\}$ as follows:
\begin{align}
\zeta_i := \pi_{\ell} \mathbb E [\mathbf x_i] +
\pi_{\mathcal S} \mathbb E [\mathbf x_i \mid \mathbf x_{\mathcal N} \geq C^*_{\mathcal N}], \ i\in\mathcal N.
\label{eq:alloc2}
\end{align}
\end{allocation}

In the next theorem, we prove that Allocation~\ref{alloc:2} provides a
sharing mechanism of the expected daily storage cost of a coalition of
agents that is in the core of the cooperative game.

\begin{theorem} \label{th:alloc2}
The cost allocation $\{\zeta_i: i\in\mathcal N\}$ defined in
Allocation~\ref{alloc:2} belongs to the core of the cost sharing
cooperative game $(\mathcal N,v)$.
\end{theorem}
\paragraph*{Proof} See appendix.

\subsubsection{Sharing of Realized Cost}
Based on the above results, the consumers can invest on joint storage and they will make savings for long term.
But the cost allocation $\zeta_i$ defined by (\ref{eq:alloc2}) is in expectation. The realized allocation will be different due to the randomness of the daily consumption.
Here we develop a daily cost allocation for the $k$-th day as
\begin{equation}
\rho_i^k=\beta_i \pi^k_\mathcal N,
\end{equation}
where $\pi^k_\mathcal N$ is the realized cost for the grand coalition on the $k$-th day and $\beta_i=\frac{\zeta_i}{\sum_{i=1}^{N} \zeta_i}$.

As $\sum_{i=1}^{N} \beta_i=1$, $\sum_{i=1}^{N}\rho_i^k=\pi^k_\mathcal N$ and the cost allocation is budget balanced. Also using strong law of large numbers, $\frac{1}{K}\sum_{k=1}^{K} \rho^k_i\rightarrow\zeta_i$ as $K\rightarrow \infty$ and the realized allocation is strongly consistent with the fixed allocation $\zeta_i$.
\section{Benefit of Cooperation}
\subsection{Scenario I}
The benefit of cooperation by joint operation of storage reflected in the total reduction of cost is given by
\begin{align}
\sum_{i\in\mathcal S} & J_i - J_S = \pi_h (\sum_{i\in\mathcal S}(x_{i} - C_{i})^+ - (x_{\mathcal S} - C_{\mathcal S})^+ ) +\nonumber\\
&\pi_{\ell}(\sum_{i\in\mathcal S}\min\{C_{i},x_{i}\}-
\min\{C_{\mathcal S},x_{\mathcal S}\}),
\end{align}
where the reduction for individual agent with cost allocation (\ref{eq:alloc1}) is
\begin{align}
J_i - \zeta_i  :=
\left\{
\begin{array}{ll}
\pi_{\delta}(C_i-x_i)^+, &
\mathrm{if} \ x_{\mathcal N} \geq C_{\mathcal N}\\
\pi_{\delta}(x_i-C_i)^+, &
\mathrm{if} \ x_{\mathcal N} < C_{\mathcal N}
\end{array}
\right.
\end{align}

\subsection{Scenario II}
The benefit of cooperation given by the reduction in the expected cost that
the coalition $\mathcal S$ obtains by jointly acquiring and exploiting the
storage is
\begin{align}
\sum_{i\in\mathcal S} & J_i^* - J_{\mathcal S}^* = \nonumber \\
&
\pi_{\mathcal S}
\sum_{i\in\mathcal S} \mathbb E [\mathbf x_i \mid \mathbf x_i \geq C_i^*] -
\pi_{\mathcal S}
\mathbb E [\mathbf x_{\mathcal S} \mid
      \mathbf x_{\mathcal S} \geq C_{\mathcal S}^*],
\end{align}
and the reduction in expected cost of each participant assuming that the
expected cost of the coalition is split using cost allocation
(\ref{eq:alloc2}) is
\begin{align}
J_i^* - \zeta_i =
\pi_{\mathcal S}
 \mathbb E [\mathbf x_i \mid \mathbf x_i \geq C_i^*] -
\pi_{\mathcal S}
\mathbb E [\mathbf x_i \mid
      \mathbf x_{\mathcal S} \geq C_{\mathcal S}^*].
\end{align}

\section{Case Study}
\label{sec:case}
We develop a case study to illustrate our results.
For this case study, we used data from the Pecan St project \cite{Pecan}.
We consider a two-period ToU tariff with $\pi_h=55$\textcent/KWh,
and $\pi_{\ell}=20$\textcent/KWh. Electricity storage is currently expensive. The amortized cost of Tesla's Powerwall Lithium-ion battery is around 25\textcent/KWh
per day. But storage prize is projected to reduce by $30\%$ by 2020 \cite{ramez2015}. Keeping in mind this projection, we consider
$\pi_{\mathcal S}=15$\textcent/KWh.

\begin{figure}
\centering
\includegraphics[width=.90\columnwidth]{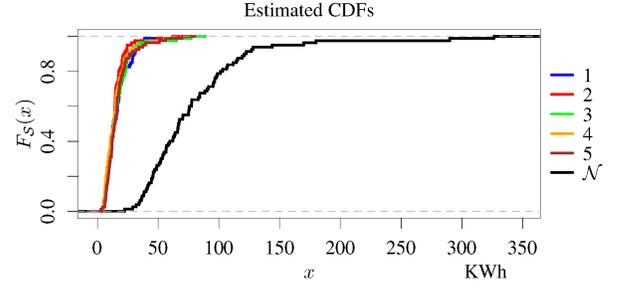}
\caption{Estimated CDFs of the peak consumption of the
five households and their aggregated consumption}
\label{fig:cdfs}
\end{figure}

A group of five household decide to join to acquire storage. Using historical
data of 2016, we estimate the individual CDFs of their daily peak
consumptions and the CDF of the daily joint peak consumption. 
Peak consumption period in Texas corresponds to 
non-holidays and non-weekends from 7h to 23h. The estimated
CDFs for peak consumption are depicted in Figure~\ref{fig:cdfs}. 
From this figure, we can see that
the shape of the CDFs are quite similar for the five households. The correlation coefficients of these five households are given in Table~\ref{tab:xcorr}. Although the shape of the CDFs are very similar,
the peak consumptions are not completely dependent. This means
that there is room for reduction in cost by making a coalition. The optimal investments in storage for the five households and for the grand coalition are given in Table~\ref{tab:res}. Also in
this table, we show the allocation of the expected storage cost
given by (\ref{eq:alloc2}). The reduction in cost for the consumers coalition
is about 5\%, however those with less correlation with the other, have a
larger reduction. Consumers 3 and 4 have cost reductions higher than 7\%, while
consumer 1, whose consumption is more correlated with the other, have about 
2.4\% of cost reduction.

\begin{table}
\caption{Correlation coefficients for the five households}
\label{tab:xcorr}
\centering
\begin{tabular}{cccccc}
\hline
& 1 & 2 & 3 & 4 & 5 \\
\hline
1 & 1.000000 & 0.363586 & 0.297733  & 0.292073 & 0.486665 \\
2 & 0.363586 & 1.000000 & 0.132320  & 0.453056 & 0.157210 \\
3 & 0.297733 & 0.132320 & 1.000000  & 0.085868 & 0.365212 \\
4 & 0.292073 & 0.453056 & 0.085869  & 1.000000 &-0.056696 \\
5 & 0.486665 & 0.157210 & 0.365212  &-0.056696 & 1.000000 \\
\hline
\end{tabular}
\end{table}

\begin{table}
\caption{Optimal storage capacity investments (in KWh),
minimal expected storage cost (in \$)
and expected cost allocation of the grand coalition (in \$)}
\label{tab:res}
\centering
\begin{tabular}{ccccccc}
\hline
& $1$ & $2$ & $3$ & $4$ & $5$ & $\mathcal N$ \\
\hline
$C_{i}^*$ &
22.98 & 14.09 & 12.64 & 13.21 & 29.82 & 95.58 \\
$J_{i}^*$ &
899.76 & 579.79 & 600.88 & 525.51 & 1189.41 & 3604.13 \\
$\zeta_i$ &
882.45 & 543.10 & 550.02 & 488.20 & 1140.35 & 3604.13 \\
\hline
\end{tabular}
\end{table}

Now, we assume that the five households buy storage independently and
then decide to cooperate by sharing their storage to reduce the realized
cost. This corresponds to Scenario I. For simplicity of computation and comparison with scenario II, we consider $\pi_i=\pi_S$ for all $i$. The realized cost is allocated
using (\ref{eq:alloc1}). In Table~\ref{tab:ralloc}, we show the allocation
of the realized aggregated cost for the ten first days of 2016, assuming
that the households have storage capacities $\{C_i^*:i\in\mathcal N\}$.

\begin{table}
\caption{Allocation of the realized cost for Scenario I for the
first ten days of the year (in \$)}
\label{tab:ralloc}
\centering
\begin{tabular}{cccccc}
\hline
Day & $\xi_1$ & $\xi_2$ & $\xi_3$ & $\xi_4$ & $\xi_5$ \\
\hline
1  & 492.66 & 612.83 & 436.88 & 549.61 & 904.69 \\
2  & 464.89 & 624.96 & 343.61 & 567.21 & 947.27 \\
3  & 541.21 & 482.61 & 299.84 & 541.40 & 820.46 \\
4  & 675.74 & 373.95 & 377.64 & 418.01 & 734.10 \\
5  & 761.41 & 403.49 & 405.52 & 371.64 & 799.23 \\
6  & 646.05 & 516.53 & 404.89 & 573.17 & 812.54 \\
7  & 654.47 & 760.99 & 387.80 & 536.92 & 797.46 \\
8  & 583.59 & 411.25 & 533.00 & 455.56 & 831.97 \\
9  & 640.46 & 394.04 & 482.85 & 483.24 & 787.20 \\
10 & 604.49 & 446.14 & 475.46 & 310.22 & 791.60 \\
\hline
\end{tabular}
\end{table}

\begin{figure}
\centering
\includegraphics[width=.90\columnwidth]{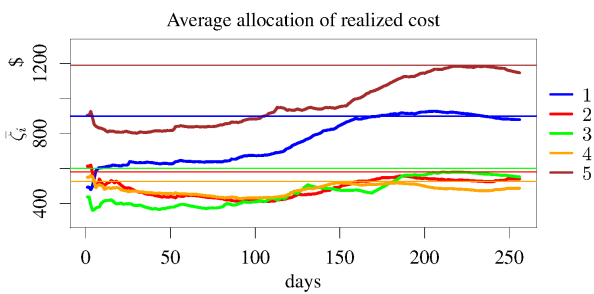}
\caption{Average allocation of the realized storage cost}
\label{fig:avrcos}
\end{figure}

Finally, in Figure~\ref{fig:avrcos}, we depict the evolution of
the average allocation of the realized cost of storage to each household
for the 2016 year. The average allocation for $D$ days is given by
\begin{align}
\bar\xi_i(D) = \frac{1}{D} \sum_{i=1}^D \xi_i, \ i\in\mathcal N,
\end{align}
where $D$ is the number of days. The average cost allocation is
compared to the optimal expected costs $J_i^*$. Assuming stationarity
of the peak consumptions random variables, the expected allocation
converge to some values
$\xi_i^{\infty} = \lim_{D\rightarrow\infty} \bar \xi_i(D) \leq J_i^*$
for $i\in\mathcal N$, as it is shown in Figure~\ref{fig:avrcos}.

\section{Conclusions}
\label{sec:concl}
In this paper, we explored sharing opportunities of electricity storage
elements among a group of consumers. We used cooperative game theory as a tool
for modeling.  Our results prove that cooperation is beneficial for agents that
either already have storage capacity or want to acquire storage capacity. In
the first scenario, the different agents only need the infrastructure to share
their storage devices. In such a case the operative scheme is really simple,
because each agent only has to storage at off-peak periods as much as possible
energy that they will consume during peak periods. At the end of the day, the
realized cost is shared among the participants. In the second scenario, the
coalition members can take an optimal decision about how much capacity they
jointly acquire by minimizing the expected daily storage cost. We showed that
the cooperative games in both the cases are balanced. We also developed
allocation rules with analytical formulas in both the cases. Thus, our results
suggest that sharing of storage in a cooperative way is very much useful for
all the agents and the society.

% References
\bibliographystyle{IEEEtran} % Name of the bibliographic style
% Other styles abbrv, acm, alpha, apalike, ieeetr, plain, siam, unsrt.
% Some publishers provide thir own styles.
\bibliography{IEEEabrv,storage} %File name with the bibliography in bibtex format.

\appendix

\section{Proofs}

\subsection{Proof of Theorem~\ref{th:subadd1}}

We shall prove that $J$ defined by (\ref{eq:rcost}) is a subadditive function.
For any nonnegative real numbers $x_{\mathcal S}$, $x_{\mathcal T}$,
$C_{\mathcal S}$, $C_{\mathcal T}$, we define
$J_{\mathcal S}=J(x_{\mathcal S},C_{\mathcal S})$,
$J_{\mathcal T}=J(x_{\mathcal T},C_{\mathcal T})$,
$J_{\mathcal S \cup \mathcal T}=
J(x_{\mathcal S}+x_{\mathcal T},C_{\mathcal S}+C_{\mathcal T})$, then
\begin{align*}
J_{\mathcal S} &=
\sum_{i\in\mathcal S}\pi_i C_i + \pi_h (x_{\mathcal S} - C_{\mathcal S})^+ +
\pi_{\ell}\min\{C_{\mathcal S},x_{\mathcal S}\}, \\
J_{\mathcal T} &=
\sum_{i\in\mathcal T}\pi_i C_i + \pi_h (x_{\mathcal T} - C_{\mathcal T})^+ +
\pi_{\ell}\min\{C_{\mathcal T},x_{\mathcal T}\}, \\
J_{\mathcal S \cup \mathcal T} &=
\sum_{i\in\mathcal S \cup \mathcal T} \pi_i C_i +
\pi_h (x_{\mathcal S} + x_{\mathcal T} - C_{\mathcal S} - C_{\mathcal T})^+ + \\
& \qquad
\pi_{\ell}\{C_{\mathcal S}+C_{\mathcal T},x_{\mathcal S}+x_{\mathcal T}\}.
\end{align*}

We can distinguish four cases\footnote{Since $x_{\mathcal S}$,
$x_{\mathcal T}$, $C_{\mathcal S}$ and, $C_{\mathcal T}$
are arbitrary nonnegative real numbers, any other possible
case can be easily recast as one of these four cases by interchanging
$\mathcal S$ and $\mathcal T$.}:
(a) $x_{\mathcal S} \geq C_{\mathcal S}$ and
$x_{\mathcal T} \geq C_{\mathcal T}$,
(b) $x_{\mathcal S} \geq C_{\mathcal S}$,
$x_{\mathcal T} < C_{\mathcal T}$ and
$x_{\mathcal S}+x_{\mathcal T} \geq C_{\mathcal S} + C_{\mathcal T}$,
(c) $x_{\mathcal S} \geq C_{\mathcal S}$,
$x_{\mathcal T} < C_{\mathcal T}$ and
$x_{\mathcal S}+x_{\mathcal T} < C_{\mathcal S} + C_{\mathcal T}$, and
(d) $x_{\mathcal S} < C_{\mathcal S}$ and
$x_{\mathcal T} < C_{\mathcal T}$.
Using simple algebra it is easy to see that for all of these cases,
$J_{\mathcal S \cup \mathcal T} \leq J_{\mathcal S}+J_{\mathcal T}$ or
equivalently,
\begin{align}
J(x_{\mathcal S}+x_{\mathcal T},C_{\mathcal S}+C_{\mathcal T})
\leq
J(x_{\mathcal S},C_{\mathcal S}) + J(x_{\mathcal T},C_{\mathcal T}),
\end{align}
and this proves subadditivity of $J$. Since the storage cost function
$u(\mathcal S)=J(x_{\mathcal S},C_{\mathcal S})$, the cost sharing
cooperative game $(\mathcal N,u)$ is subadditive.
\qed

\subsection{Proof of Theorem~\ref{th:bal1}}

We notice that the function $J$ is positive
homogeneous, i.e, for any $\alpha \geq 0$,
\(
J(\alpha x_{\mathcal S},\alpha C_{\mathcal S}) =
\alpha J(x_{\mathcal S},C_{\mathcal S}).
\)
$J$ is also
subadditive as per Theorem~\ref{th:subadd1}. Thus
for any arbitrary balanced map $ \alpha: 2^\mathcal{N}\rightarrow[0,1]$
\begin{align*}
\sum_{\mathcal S \in 2^{\mathcal N}} & \alpha(\mathcal S)u(\mathcal S) \\
&= \sum_{\mathcal S \in 2^{\mathcal N}} \alpha(\mathcal S)
J (x_{\mathcal S},C_{\mathcal S}) \\
&= \sum_{\mathcal S \in 2^{\mathcal N}}
J( \alpha(\mathcal S)x_{\mathcal S},
   \alpha(\mathcal S) C_{\mathcal S}) \ \text{[positive homogeneity]}\\
&\geq J( \sum_{\mathcal S \in 2^{\mathcal N}}
        \alpha(\mathcal S) x_{\mathcal S},
        \sum_{\mathcal S \in 2^{\mathcal N}}
        \alpha(\mathcal S) C_{\mathcal S}) \ \text{[subadditivity]} \\
&= J( \sum_{i\in\mathcal N} \sum_{\mathcal S \in 2^{\mathcal N}}
        \alpha(\mathcal S) \mathbf 1_{\mathcal S}(i)x_{\mathcal S},
        \sum_{i\in\mathcal N} \sum_{\mathcal S \in 2^{\mathcal N}}
        \alpha(\mathcal S) \mathbf 1_{\mathcal S}(i)C_{\mathcal S})
        \\
&= J(x_{\mathcal N},C_{\mathcal N}) = u(\mathcal N).
\end{align*}
and this proves that the cost sharing game $(\mathcal N, u)$ is
balanced.
\qed

\subsection{Proof of Theorem~\ref{th:alloc1}}
We begin by proving that the cost allocation (\ref{eq:alloc1}) is
an imputation, \emph{i.e.} $\xi \in \mathcal I$. An imputation is a cost
allocation satisfying budget balance and individual rationality.

If $x_{\mathcal N} \geq C_{\mathcal N}$:
\begin{align*}
%u(\mathcal N) &=
%\sum_{i\in\mathcal N} \pi_i C_i + \pi_h(x_{\mathcal N}-C_{\mathcal N}) +
%\pi_{\ell} C_{\mathcal N} \\
\sum_{i\in\mathcal N} \xi_i &=
\sum_{i\in\mathcal N} \pi_i C_i + \pi_h(x_{\mathcal N}-C_{\mathcal N}) +
\pi_{\ell} C_{\mathcal N} = u(\mathcal N).
\end{align*}

If $x_{\mathcal N} < C_{\mathcal N}$:
\begin{align*}
%u(\mathcal N) &=
%\sum_{i\in\mathcal N} \pi_i C_i + \pi_{\ell} x_{\mathcal N} \\
%\sum_{i\in\mathcal N} \xi_i &=
\sum_{i\in\mathcal N} \pi_i C_i + \pi_{\ell} x_{\mathcal N}
= u(\mathcal N).
\end{align*}
Thus, $\sum_{i\in\mathcal N} \xi_i = u(\mathcal N)$ and the cost
allocation $\{\xi_i:i\in\mathcal N\}$ satisfies budget balance.

The individual cost is:
\begin{align*}
u(\{i\}) &= \left\{
\begin{array}{ll}
\pi_i C_i + \pi_h(x_i-C_i) + \pi_{\ell} C_i & x_i \geq C_i \\
\pi_i C_i + \pi_{\ell} x_i & x_i < C_i
\end{array}
\right.
\end{align*}

If $x_{\mathcal N} \geq C_{\mathcal N}$:
\begin{align*}
\xi_i &= \pi_i C_i+ \pi_h(x_i-C_i)+\pi_{\ell} C_i \\
&= \pi_i C_i +\pi_{\ell}x_i - \pi_{\delta}(C_i-x_i) \\
&= u(\{i\})-\pi_{\delta}(C_i-x_i)^+.
\end{align*}

If $x_{\mathcal N} < C_{\mathcal N}$:
\begin{align*}
\xi_i &= \pi_i C_i+ \pi_{\ell} x_i \\
&= u(\{i\}) - \pi_{\delta}(x_i-C_i)^+.
\end{align*}
Thus, $\xi_i \leq v(\{i\})$ for all $i\in\mathcal N$, and the cost allocation
$\xi$ is individually rational. Since it is also budget balanced,
it is an imputation, \emph{i.e.} $\xi \in \mathcal I$.

Finally, to prove that the cost allocation $\xi$ belongs to the core of the
cooperative game, we have to prove that
$\sum_{i\in\mathcal S}\xi_i \leq u(\mathcal S)$ for any
coalition $\mathcal S \subseteq \mathcal N$.

If $x_{\mathcal N} \geq C_{\mathcal N}$:
\begin{align*}
\sum_{i\in\mathcal S} \xi_i &=
\sum_{i\in\mathcal S} \pi_i C_i+
\pi_h(x_{\mathcal S}-C_{\mathcal S})+\pi_{\ell} C_{\mathcal S} \\
&= \sum_{i\in\mathcal S} \pi_i C_i +\pi_{\ell}x_{\mathcal S} -
\pi_{\delta}(C_{\mathcal S}-x_{\mathcal S}) \\
&= u({\mathcal S})-\pi_{\delta}(C_{\mathcal S}-x_{\mathcal S})^+.
\end{align*}

If $x_{\mathcal N} < C_{\mathcal N}$:
\begin{align*}
\sum_{i\in\mathcal S}\xi_i &=
\sum_{i\in\mathcal S}\pi_{\mathcal S} C_{\mathcal S}+
\pi_{\ell} x_{\mathcal S} \\
&= u({\mathcal S}) - \pi_{\delta}(x_{\mathcal S}-C_{\mathcal S})^+.
\end{align*}
Thus, $\sum_{i\in\mathcal S} \xi_i \leq u(\mathcal S)$ for any
$\mathcal S \subseteq \mathcal N$ and the cost allocation $\xi$ is
in the core of the cooperative game $(\mathcal N, u)$.
\qed

\subsection{Proof of Theorem~\ref{th:subadd2}}
Let $\mathcal S$ and $\mathcal T$ two arbitrary nonempty disjoint coalitions,
\emph{i.e.} $\mathcal S,\mathcal T \subseteq \mathcal N$ such that
$\mathcal S \cap \mathcal T = \emptyset$. Define
\begin{align}\label{eq:defphi}
\Phi(\mathbf x_{\mathcal S}) &=
\min_{C_{\mathcal S}\geq 0} \mathbb E J(C_{\mathcal S},\mathbf x_{\mathcal S}).
\end{align}
We shall prove that $\Phi(\mathbf x_{\mathcal S})$ is a subbadditive function.

From the definition of $J$ given in (\ref{eq:rcost}),
\begin{align*}
J(\mathbf x_{\mathcal S},C^*_{\mathcal S})+
J(\mathbf x_{\mathcal T},C^*_{\mathcal T}) \geq
J(\mathbf x_{\mathcal S}+\mathbf x_{\mathcal T},C^*_{\mathcal S}+C^*_{\mathcal T}).
\end{align*}
Taking expectations on both sides,
\begin{align*}
\Phi(\mathbf x_{\mathcal S}) +
\Phi(\mathbf x_{\mathcal T})
&\geq
\mathbb E J(\mathbf x_{\mathcal S}+\mathbf x_{\mathcal T},C^*_{\mathcal S}+C^*_{\mathcal T})\\
&\geq \min_{C\geq 0}
\mathbb E J(\mathbf x_{\mathcal S}+\mathbf x_{\mathcal T},C) \\
&= \Phi(\mathbf x_{\mathcal S},\mathbf x_{\mathcal T}),
\end{align*}
and this proves subadditivity of $\Phi$.

Subadditivity of the cost sharing cooperative game $(\mathcal N,v)$ is
a consequence of the subadditivity of $\Phi$ because
$v(\mathcal S) = \Phi(\mathbf x_{\mathcal S})$ for any
$\mathcal S \subseteq \mathcal N$.
\qed

\subsection{Proof of Theorem~\ref{th:bal2}}

First, we prove that the function
$\Phi$ defined by (\ref{eq:defphi}) is positive homogeneous.
Observe that if a random variable $z$ has CDF $F$,
then the scaled random variable $\alpha z$
with $\alpha>0$ has CDF:
\(
F_{\alpha}(\theta) = {\mathbb P}\{\alpha z \leq \theta\} = F(\theta/\alpha).
\)
Then, for any $\alpha\geq 0$ and $\gamma\in [0,1]$,
$\gamma = F(C)$ if and only if $\gamma = F_{\alpha}(\alpha C)$.
This means that if $C_{\mathcal S}$ is such that
$\Phi(\mathbf x_{\mathcal S})=\mathbb E J(\mathbf x_{S},C_{S}^*)$, then
$\Phi(\alpha \mathbf x_{\mathcal S})=\mathbb E J(\alpha \mathbf x_{S},\alpha C_{S}^*)$.

For any $\alpha \geq 0$, and from the definition of the daily
storage cost $J$ (\ref{eq:rcost}),
\(
J(\alpha \mathbf x_{\mathcal S},\alpha C^*_{\mathcal S}) =
\alpha J(\mathbf x_{\mathcal S},C^*_{\mathcal S}).
\)
Taking expectations on both sides,
\(
\Phi(\alpha \mathbf x_{\mathcal S}) = \alpha \Phi(\mathbf x_{\mathcal S}),
\)
and this proves positive homogeneity of $\Phi$.

Now, balancedness of the cost sharing cooperative game
$(\mathcal N,v)$ is a consequence of the properties of
function $\Phi$
\begin{align*}
\sum_{\mathcal S \in 2^{\mathcal N}} \alpha(\mathcal S) v(\mathcal S)
&= \sum_{\mathcal S \in 2^{\mathcal N}}
   \alpha(\mathcal S) \Phi(\mathbf x_{\mathcal S})\\
&= \sum_{\mathcal S \in 2^{\mathcal N}}
    \Phi(\alpha(\mathcal S)\mathbf x_{\mathcal S}) \ \text{[positive homogeneity]}\\
&\geq
\Phi(\sum_{\mathcal S \in 2^{\mathcal N}}
      \alpha(\mathcal S)\mathbf x_{\mathcal S}) \ \text{[subadditivity]} \\
&= \Phi(\sum_{i\in\mathcal N} \sum_{\mathcal S \in 2^{\mathcal N}}
        \alpha(\mathcal S) \mathbf 1_{\mathcal S}(i)\mathbf x_{\mathcal S})\\
&= \Phi(\mathbf x_{\mathcal N})= v(\mathcal N).
\end{align*}

\subsection{Proof of Theorem~\ref{th:alloc2}}
We begin by proving that the cost allocation given by (\ref{eq:alloc1})
satisfies budget balance,
\begin{align*}
%v(\mathcal N) &=
%\sum_{i\in\mathcal N} \pi_{\ell} \mathbb E[\mathbf x_{\mathcal N}]
%+ \pi_h \mathbb E[\mathbf x_{\mathcal N}\mid \mathbf x_{\mathcal N} \geq C^*_{\mathcal N}] \\
\sum_{i\in\mathcal N} \zeta_i
&=
\sum_{i\in\mathcal N}
\pi_{\ell} \mathbb E [\mathbf x_i] +
\sum_{i\in\mathcal N}
\pi_{\mathcal S} \mathbb E [\mathbf x_i \mid \mathbf x_{\mathcal N} \geq C^*_{\mathcal N}] \\
&=
\pi_{\ell} \mathbb E \left[ \sum_{i\in\mathcal N} \mathbf x_i \right] +
\pi_{\mathcal S} \mathbb E \left[ \sum_{i\in\mathcal N} \mathbf x_i \mid
\mathbf x_{\mathcal N} \geq C^*_{\mathcal N}\right] \\
&=
\pi_{\ell} \mathbb E[\mathbf x_{\mathcal N}]
+ \pi_h \mathbb E[\mathbf x_{\mathcal N}\mid \mathbf x_{\mathcal N} \geq C^*_{\mathcal N}] \\
&= v(\mathcal N).
\end{align*}
The cost allocation is in the core if we prove that $v(\mathcal S) \geq \sum_{i\in\mathcal S}\zeta_i$
for any coalition $\mathcal S\subset \mathcal N$. Please note that individual
rationality is included in the previous condition.

The storage cost for a coalition $\mathcal S\subset\mathcal N$ is
\begin{align*}
v(\mathcal S)
&=
\pi_{\ell} \mathbb E[\mathbf x_{\mathcal S}] +
\pi_{\mathcal S} \mathbb E[\mathbf x_{\mathcal S}\mid \mathbf x_{\mathcal S}\geq C_{\mathcal S}^*] \\
&=
\pi_{\mathcal S} C_{\mathcal S}^* +
\pi_h \mathbb E[(\mathbf x_{\mathcal S}-C_{\mathcal S}^*)^+] +
\pi_{\ell} \mathbb E[\min\{C_{\mathcal S}^*,\mathbf x_{\mathcal S}\}].
\end{align*}
Note that
\begin{align*}
\pi_h (\mathbf x_{\mathcal S}-C^*_{\mathcal S})^{+} +
\pi_{\ell} \min\{C^*_{\mathcal S},\mathbf x_{\mathcal S}\} \geq
\pi_h (\mathbf x_{\mathcal S}-C^*_{\mathcal S}) + \pi_{\ell} C^*_{\mathcal S}.
\end{align*}
and therefore,
\begin{align*}
\pi_{\mathcal S} C^*_{\mathcal S}
+ \pi_h\mathbb{E}[(\mathbf x_{\mathcal S}-C^*_{\mathcal S})^{+}]
+ \pi_{\ell} \mathbb{E}[\min\{C^*_{\mathcal S},\mathbf x_{\mathcal S}\}] \qquad \\
\geq \pi_{\mathcal S} C^*_{\mathcal S} +
\pi_h \mathbb{E}[(\mathbf x_{\mathcal S}-C^*_{\mathcal S})]
+ \pi_{\ell} C^*_{\mathcal S}.
\end{align*}

Let us define the sets
$\mathcal A_+ =
\{\mathbf x_{\mathcal N}\in\mathbb R_+\mid \mathbf x_{\mathcal N} \geq C_{\mathcal N}$\},
$\mathcal A_- = \mathbb R_+ \backslash \mathcal A_+$, and the
auxiliary function $\psi(\mathbf x_{\mathcal N})$ as follows
\begin{align*}
\psi(\mathbf x_{\mathcal N})=
\left\{
\begin{array}{cc}
\pi_h \ \text{if} \ \mathbf x_{\mathcal N} \in \mathcal A_+ \\
\pi_{\ell} \ \text{if} \ \mathbf x_{\mathcal N} \in \mathcal A_-
\end{array}
\right.
\end{align*}

Let $F(\mathbf x_{\mathcal S},\mathbf x_{\mathcal N})$ be the joint distribution function
of the peak consumptions $(\mathbf x_{\mathcal S}, \mathbf x_{\mathcal N})$, then
\begin{align*}
\mathbb{E} & [\psi(\mathbf x_{\mathcal N})(\mathbf x_{\mathcal N}-C^*_{\mathcal N})]\\
&=
\pi_{\ell} \int_{\mathbb R_+} \int_{\mathcal A_-}
(\mathbf x_{\mathcal S}-C^0_{\mathcal S}) dF(\mathbf x_{\mathcal S},\mathbf x_{\mathcal N}) + \\
& \qquad
\pi_h \int_{\mathbb R_+} \int_{\mathcal A_+}
(\mathbf x_{\mathcal S}-C^*_{\mathcal S}) dF(\mathbf x_{\mathcal S},\mathbf x_{\mathcal N}) \\
& \leq
\pi_h\int_{\mathbb R_+} \int_{\mathcal A_+ \cup \mathcal A_-}
(\mathbf x_{\mathcal S}-C^*_{\mathcal S}) dF(\mathbf x_{\mathcal S},\mathbf x_{\mathcal N}) \\
&=\pi_h\int_{\mathbb R_+}
(\mathbf x_{\mathcal S}-C^*_{\mathcal S}) dF(\mathbf x_{\mathcal S},\mathbf x_{\mathcal N}) \\
&= \mathbb E [(\mathbf x_{\mathcal S}-C^*_{\mathcal S}],
\end{align*}
and consequently,
\begin{align*}
\pi_{\mathcal S} C^*_{\mathcal S} &+
\pi_h \mathbb{E}[(\mathbf x_{\mathcal S}-C^*_{\mathcal S})] +
\pi_{\ell} C^*_{\mathcal S} \\
&\geq
\pi_{\mathcal S} C^*_{\mathcal S} +
\mathbb{E}[\pi_{\alpha}(\mathbf x_{\mathcal S}-C^*_{\mathcal S})] +
\pi_{\ell} C^*_{\mathcal S}
\end{align*}

Now, we prove that the right hand side of the previous expression equals
$\sum_{i\in\mathcal S} \zeta_i$
\begin{align*}
\pi_{\mathcal S} & C^*_{\mathcal S} +
\mathbb{E}[\psi(\mathbf x_{\mathcal N})(\mathbf x_{\mathcal S}-C^*_{\mathcal S})] +
\pi_{\ell} C^*_{\mathcal S} \\
&=\pi_{\mathcal S} C^*_{\mathcal S} +
\int_{R_+} \int_{R_+}
\psi(\mathbf x_{\mathcal N})(\mathbf x_i-C^0_i)dF(\mathbf x_{\mathcal S},\mathbf x_{\mathcal N}) +
\pi_{\ell} C^*_{\mathcal S} \\
%&=\pi_{\mathcal S} C^*_{\mathcal S}+
%\pi_{\ell} \int_{\mathbb R_+} \int_{\mathcal A_-}
%(\mathbf x_{\mathcal S}-C^*_{\mathcal S}) dF(\mathbf x_{\mathcal S},\mathbf x_{\mathcal N}) +\\
%&\qquad
%\pi_h \int_{\mathbb R_+} \int_{\mathcal A_+}
%(\mathbf x_{\mathcal S}-C^*_{\mathcal S}) dF(\mathbf x_{\mathcal S},\mathbf x_{\mathcal N}) +
%\pi_{\ell} C^*_{\mathcal S} \\
&=\pi_{\mathcal S} C^*_{\mathcal S} +
\pi_{\ell} \int_{\mathbb R_+} \int_{\mathcal A_-\cup\mathcal A_+}
(\mathbf x_{\mathcal S}-C^*_{\mathcal S}) dF(\mathbf x_{\mathcal S},\mathbf x_{\mathcal N}) + \\
&\qquad
(\pi_h-\pi_{\ell}) \int_{\mathbb R_+} \int_{\mathcal A_+}
(\mathbf x_{\mathcal S}-C^*_{\mathcal S}) dF(\mathbf x_{\mathcal S},\mathbf x_{\mathcal N})
+ \pi_{\ell} C^*_{\mathcal S} \\
&=\pi_{\mathcal S} C^*_{\mathcal S} +
\pi_{\delta} \int_{\mathbb R_+} \int_{\mathcal A_+}
(\mathbf x_{\mathcal S}-C^i_{\mathcal S}) dF(\mathbf x_{\mathcal S},\mathbf x_{\mathcal N})
+ \pi_{\ell} \mathbb{E}[\mathbf x_i] \\
&=\pi_{\mathcal S} C^*_{\mathcal S} +
\frac{\pi_{\mathcal S}}{1-\gamma_{\mathcal S}}  \int_{\mathbb R_+} \int_{\mathcal A_+}
(\mathbf x_{\mathcal S}-C^*_{\mathcal S}) dF(\mathbf x_{\mathcal S},\mathbf x_{\mathcal N}) +
   \pi_{\ell} \mathbb{E}[\mathbf x_i] \\
&= \pi_{\mathcal S}\frac{1}{\mathbb P\{\mathbf x_{\mathcal N} \geq C_{\mathcal N}\}}
   \int_{\mathbb R_+} \int_{\mathcal A_+}
   \mathbf x_{\mathcal S}dF(\mathbf x_{\mathcal S},\mathbf x_{\mathcal N}) +
   \pi_{\ell} \mathbb{E}[\mathbf x_i] \\
%&=
%\pi_{\mathcal S} \mathbb E [\mathbf x_{\mathcal S} \mid X_{\mathcal N}\geq C^*_{\mathcal N}] +
%\pi_{\ell} \mathbb E [\mathbf x_{\mathcal S}]\\
&=\sum_{i\in\mathcal S} \zeta_i
\end{align*}

Thus, $\sum_{i\in\mathcal S} \zeta_i \leq v(\mathcal S)$ and the cost allocation
$\{\zeta_i : i \in \mathcal N\}$ is an imputation in the core.

\end{document}